\documentclass[prl,twocolumn,superscriptaddress]{revtex4-1}
\usepackage{}
\usepackage{amssymb}
\usepackage{amsfonts}
\usepackage{bbm}
\usepackage{mathrsfs}
\usepackage{graphics,graphicx,epsfig,bm,amsmath,amsthm,amssymb}
\usepackage{bm}
\usepackage{bbm}
\usepackage{longtable}
\usepackage{multirow}
\usepackage{array}
\usepackage{color}
\usepackage[usenames,dvipsnames]{xcolor}
\usepackage{booktabs}
\usepackage{verbatim}

\usepackage{float}
\usepackage[a4paper,colorlinks=true,
linkcolor=blue,citecolor=blue,
pdfauthor={ },
pdftitle={ },
pdfsubject={ },
pdfkeywords={ }]{hyperref}

\begin{document}

\title{Experimental investigation of quantum correlations in a two-qutrit spin system}

\author{Yue Fu}
\thanks{These authors contributed equally to this work.}
\affiliation{CAS Key Laboratory of Microscale Magnetic Resonance and School of Physical Sciences, University of Science and Technology of China, Hefei 230026, China}
\affiliation{CAS Center for Excellence in Quantum Information and Quantum Physics, University of Science and Technology of China, Hefei 230026, China}

\author{Wenquan Liu}
\thanks{These authors contributed equally to this work.}
\affiliation{CAS Key Laboratory of Microscale Magnetic Resonance and School of Physical Sciences, University of Science and Technology of China, Hefei 230026, China}
\affiliation{CAS Center for Excellence in Quantum Information and Quantum Physics, University of Science and Technology of China, Hefei 230026, China}

\author{Xiangyu Ye}
\affiliation{CAS Key Laboratory of Microscale Magnetic Resonance and School of Physical Sciences, University of Science and Technology of China, Hefei 230026, China}

\author{Ya Wang}
\affiliation{CAS Key Laboratory of Microscale Magnetic Resonance and School of Physical Sciences, University of Science and Technology of China, Hefei 230026, China}
\affiliation{CAS Center for Excellence in Quantum Information and Quantum Physics, University of Science and Technology of China, Hefei 230026, China}
\affiliation{Hefei National Laboratory, University of Science and Technology of China, Hefei 230088, China}

\author{Chengjie Zhang}
\affiliation{School of Physical Science and Technology, Ningbo University, Ningbo 315211, China}
\affiliation{State Key Laboratory of Precision Spectroscopy, School of Physics and Electronic Science, East China Normal University, Shanghai 200241, China}

\author{Chang-Kui Duan}
\affiliation{CAS Key Laboratory of Microscale Magnetic Resonance and School of Physical Sciences, University of Science and Technology of China, Hefei 230026, China}
\affiliation{CAS Center for Excellence in Quantum Information and Quantum Physics, University of Science and Technology of China, Hefei 230026, China}

\author{Xing Rong}
\email{xrong@ustc.edu.cn}
\affiliation{CAS Key Laboratory of Microscale Magnetic Resonance and School of Physical Sciences, University of Science and Technology of China, Hefei 230026, China}
\affiliation{CAS Center for Excellence in Quantum Information and Quantum Physics, University of Science and Technology of China, Hefei 230026, China}
\affiliation{Hefei National Laboratory, University of Science and Technology of China, Hefei 230088, China}

\author{Jiangfeng Du}
\email{djf@ustc.edu.cn}
\affiliation{CAS Key Laboratory of Microscale Magnetic Resonance and School of Physical Sciences, University of Science and Technology of China, Hefei 230026, China}
\affiliation{CAS Center for Excellence in Quantum Information and Quantum Physics, University of Science and Technology of China, Hefei 230026, China}
\affiliation{Hefei National Laboratory, University of Science and Technology of China, Hefei 230088, China}

\begin{abstract}
We report an experimental investigation of quantum correlations in a two-qutrit spin system in a single nitrogen-vacancy center in diamond at room temperatures.  Quantum entanglement between two qutrits was observed at room temperature and the existence of non-classical correlations beyond entanglement in the qutrit case has been revealed.
 Our work demonstrates the potential of the NV centers as the multi-qutrit system to execute quantum information tasks and provides a powerful experimental platform for studying fundamental physics of high-dimensional quantum systems in future.
\end{abstract}
\maketitle

Quantum correlation shed light on the most fundamental trait that distinguishes a quantum correlated system from one fully ascribed to a joint classical probability distribution and may reveal the origin of the quantum enhancement in various quantum information tasks \cite{JPhys_2016_Adesso,review_2018_De,book_2015_streltsov,review_2018_Braun}.
Quantum entanglement as a prolonged description of quantum correlation \cite{review_2009}, however, has been found cannot account for all quantum correlations, leading to the introduction of quantum discord \cite{PRL_2001_Olivier,JPA_2001_Henderson}.
Quantum discord can describe non-classical correlations even in separable states and may contribute to the quantum enhancement when these states are applied to quantum information processing (QIP) \cite{PRL_1998_Knill,review_2012_Modi,review_2018_Bera}.
In the last decades, quantum discord was studied intensively but experimental investigations have been limited to qubit systems \cite{PRA_2010_Soares,PRL_2011_Auccaise,PRA_2017_Singh,PRL_2008_Lanyon,NC_2010_Xu,NC_2013_Xu,PRL_2015_Bromley,PRA_2016_Knoll,PRB_2011_Yurishchev,PRB_2012_Rong,PRB_2013_Rong,NP_2013_Gessner,NC_2017_Abdelrahman,PRA_2014_Wood}.
Recently, researches on systems with $d$-level (${d>2}$), hereafter referred to as qudits, are emerging \cite{AQT_2019_Cozzolino,FP_2020_Wang,review_2020_Erhard}.
Quantum information processing with qudit may offer higher-dimensional Hilbert space, which results in higher efficiency and flexibility in quantum computing \cite{PRL_2019_Wang,NP_2019_Reimer}, larger channel capacity and better noise tolerance in quantum communication \cite{PRL_2019_Luo,PRL_2020_Hu} and relaxed constraints in fundamental tests of the nature \cite{PRL_2010_Vertesi}.
Currently, the experimental studies of quantum correlations in qudit systems have been focused on quantum entanglement \cite{nature_2017_Kues,Science_2018_Wang,NPJ_2020_Lu,arxiv_2021_Ding,LSA_2016_Ding} and quantum steering \cite{PRL_2018_Zeng, PRL_2019_Guo,PRL_2021_Designolle}.
Quantum discord, a more general description of the quantum correlation, remains untouched experimentally in qutrit systems.

In this paper, we report an experimental investigation of quantum correlations in a two-qutrit system. The nitrogen-vacancy (NV) center, which has a spin-1 electron spin with the complement of a spin-1 nuclear spin of the $^{14}$N \cite{PhysRep_2013_Doherty}, was utilized as a two-qutrit system to study the high dimensional quantum correlations.
The quantum correlations of two-qutrit isotropic states (also known as qutrit Werner states \cite{QIP_2013_Ye}), which are crucial in the research of quantum correlation \cite{PRA_2006_Baumgartner, review_2009, PRA_2012_Chitambar, QIP_2013_Ye}, were prepared and measured here. The features of quantum discord and quantum entanglement with different values of the state parameter are revealed. It is verified that there is a threshold under which the quantum entanglement vanishes while the quantum discord remains.
Such a threshold differs from the one in the case of a two-qubit system \cite{PRA_2006_Baumgartner, PRA_2012_Chitambar, QIP_2013_Ye, PRA_2021_Poxleitner}.

\begin{figure}\centering
	\includegraphics[width=1\columnwidth]{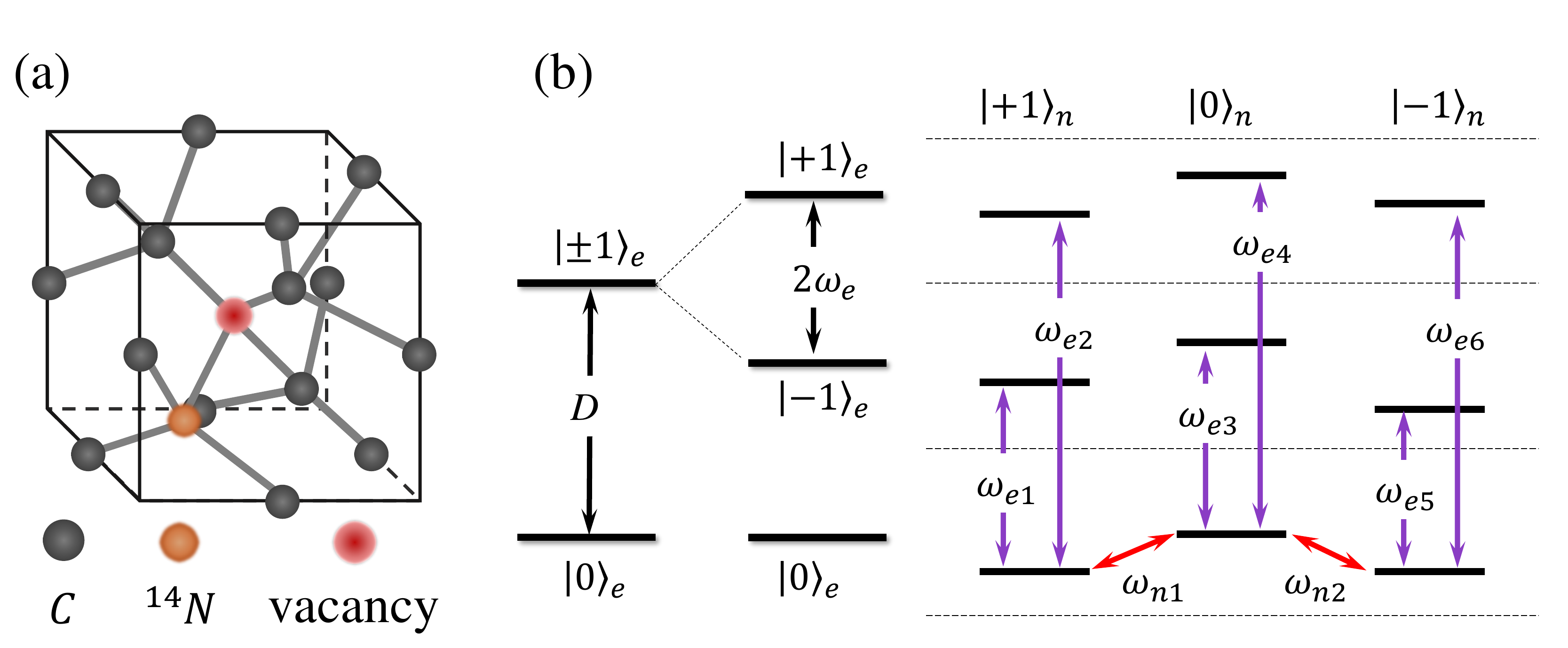}
	\caption{The two-qutrit system constructed by the NV center. (a) Schematic atomic structure of the NV center.
(b) Ground state energy levels of the NV center.
The nine different states of the electron spin and the nuclear spin constitute a two-qutrit system.
The transitions between different electron (nuclear) spin states can be steered by microwave (radio frequency) pulses indicated by purple (red) arrows.
}
	\label{Fig1}
\end{figure}

The total correlation of a bipartite system $\rho_{AB}$ quantified by the quantum mutual information \cite{JPA_2001_Henderson, PRL_2001_Olivier} is defined as
\begin{align}
	I(\rho_{AB}) = S(\rho_A)+S(\rho_B)-S(\rho_{AB}),
\end{align}
where $\rho_A$ ($\rho_B$) is the reduced density matrix of particle $A $ ($B$) and $S(\rho) = -\rm{Tr}[\rho\rm{log_2}\rho]$ is the von Neumann entropy of density matrix $\rho$.
Both classical and quantum correlations are included in $I(\rho_{AB})$.
The classical correlation, which depends on the maximum information gained by measuring one particle of the total system, is defined as
 \begin{align}
	C(\rho_{AB}) = {\rm max}_{B_j^\dagger B_j}[ S(\rho_A)-\sum_jq_jS(\rho_A^j)],
	\label{eq.2}
\end{align}
where $q_j ={ \rm Tr}[B_j\rho_{AB}B_j^\dagger]$ is the probability of obtaining result $j$ when performing Positive Operator-Valued Measure (POVM)
$\{B_j^\dagger B_j\}$ on subsystem $B$,
$\rho_A^j = {\rm Tr}_B[B_j\rho_{AB}B_j^\dagger]/q_j$ is the state of subsystem $A$ after obtaining outcome $j$.
Therefore, the difference between the total correlation and the classical correlation given by
\begin{align}
	D(\rho_{AB}) & = I(\rho_{AB})-C(\rho_{AB}) ,
	\label{eq.3}
\end{align}
is quantum correlation, which is termed quantum discord.

We focus on a family of qutrit states referred to as isotropic states, which have the form \cite{PRA_2006_Baumgartner}
\begin{align}
	\rho_{\rm iso} = \frac{(1-p)}{9}\mathbb{I}_3 \otimes \mathbb{I}_3+ p|\psi\rangle \langle\psi|,
\label{eq4}
\end{align}
where $|\psi\rangle = (|{+1,+1}\rangle+|0,0\rangle+|{-1,-1}\rangle)/\sqrt{3}$ is the maximally entangled state, $p \in [0,1]$ and $\mathbb{I}_{n}$ denotes the identity operator in $n$-dimensional Hilbert space.
Physically, an isotropic state can be regarded as a mixture of the maximally mixed state $\mathbb{I}_9/9$ and the maximally entangled state $|\psi\rangle\langle\psi|$ with parameter $p$ determining the components.
Theoretically, there is a threshold $p_c$ under which the states are separable and otherwise entangled.
The threshold varies with the dimension of the isotropic states and $p_c=1/4
$ for qutrits \cite{PRA_2006_Baumgartner, PRA_2012_Chitambar}.
However, the quantum discord remains non-vanishing for all non-zero $p$.
Thus, the isotropic states are a representative example of quantum states with zero entanglement nevertheless exhibiting non-zero quantum correlations.

\begin{figure}\centering
\includegraphics[width=1\columnwidth]{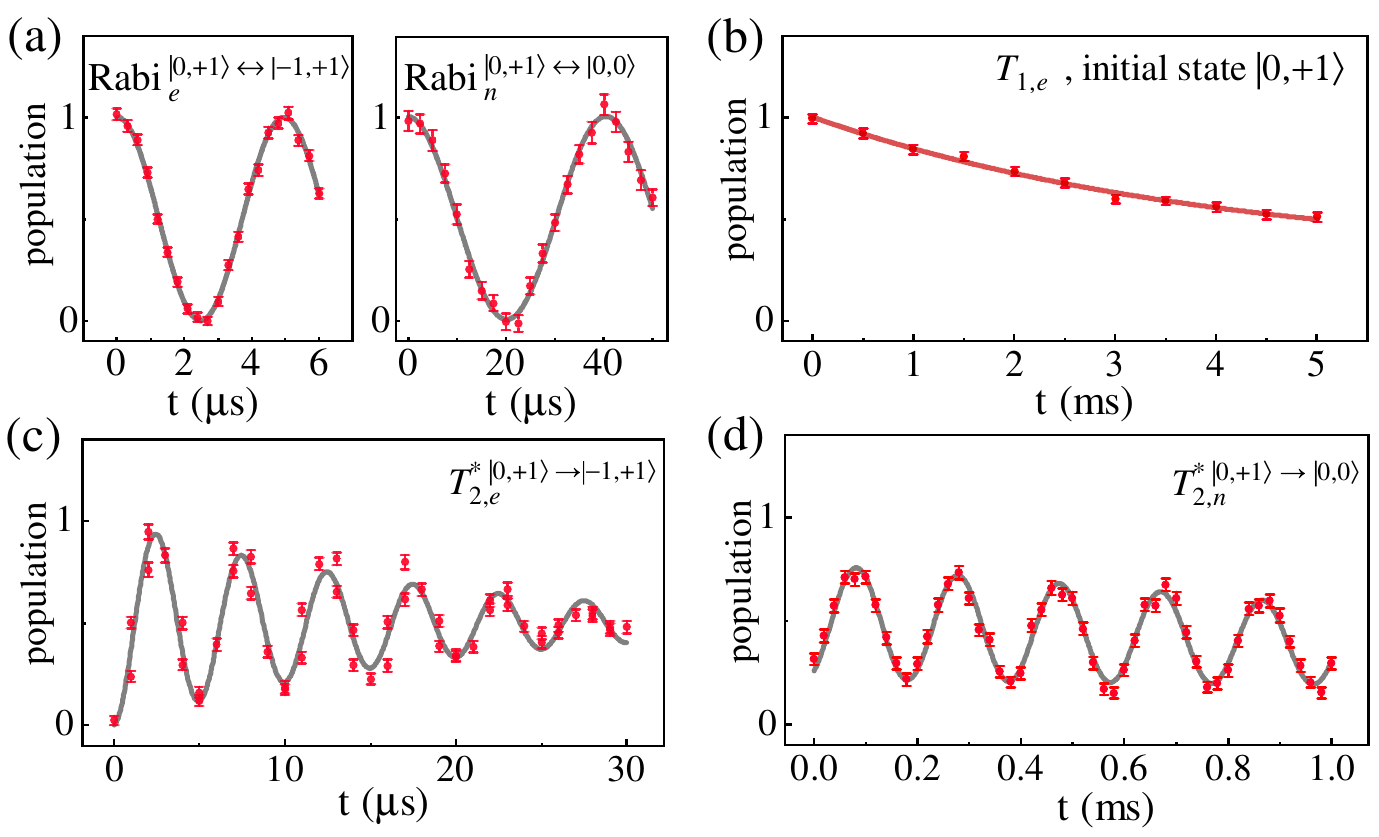}
\caption{The properties of the two-qutrit system. (a) Rabi oscillation between different electron (left panel) or nuclear (right panel) spin states. (b) Longitudinal relaxation time of the electron spin $T_{1,e}$ starting form the state $|0,{+1}\rangle$. (c)-(d) Dephasing time of the electron spin and the nuclear spin. All points with error bars are experimental data and the curves are the fitting results.
 }
\label{Fig2}
\end{figure}

A single negatively charged NV center in [100] face bulk diamond was utilized to investigate the quantum discord and quantum entanglement of a two-qutrit system.
The diamond was isotopically purified ($[^{12}C = 99.9\%]$) to enhance the dephasing time of the electron spin.
The NV center consists of a substitutional nitrogen atom adjacent to a carbon vacancy (see Fig.~\ref{Fig1}a).
When an external magnetic field is applied along the NV symmetry axis, the Hamiltonian of the NV center can be written as
\begin{align}
	H_{\rm NV} = 2\pi(DS_z^2+\omega_{e}S_z+QI_z^2+\omega_{ n}I_z+AS_zI_z),
\end{align}
where $S_z\ (I_z) $ is the spin operator of the electron (nuclear) spin, $D = 2.87$ GHz is the electronic zero-field splitting, $Q = -4.95$ MHz is the nuclear quadrupolar interaction constant, and $A = -2.16$ MHz is the hyperfine coupling constant.
$\omega_e$ ($\omega_n$) corresponds to the Zeeman frequency of the electron (nuclear) spin.
This two-qutrit system contains nine energy levels denoted as $|m_S\rangle_e|m_I\rangle_n$, with $m_S,m_I=0,\,\pm1$ representing the states of the electron and nuclear spins, respectively (see Fig.~\ref{Fig1}b).
For simplicity, $|m_S\rangle_e|m_I\rangle_n$ is hereafter labeled by $|m_S,m_I\rangle$.
The magnetic field was set to 500 G and the NV center was polarized into the state $|{0,+1}\rangle$ via applying a 532 nm laser pulse \cite{PRL_2009_Jacques}.
As shown in Fig.~\ref{Fig1}b, microwave (MW) pulses labeled by purple arrows were applied to manipulate the quantum states of the electron spin of the NV center.
For the controlling of the nuclear spins, radio frequency (RF) pulses labeled by red arrows were utilized.

The Rabi frequency of the electron (nuclear) spin was set to $\Omega_{\rm MW} = 0.2$ MHz ($\Omega_{\rm RF} = 25$ kHz) as depicted in Fig.~\ref{Fig2}a.
The relaxation times of both the electron spin and the nuclear spin were measured.
For the electron spin, the longitudinal relaxation time is $T_{1,e} = 4 \pm 1$ ms (Fig.~\ref{Fig2}b) and the dephasing time is $T_{2,e}^* = 18\pm 2\ \mu$s (Fig.~\ref{Fig2}c).
In Fig.~\ref{Fig2}d, we did not see evident decay of the oscillation amplitude during the measurement, which indicated that the dephasing time of the nuclear spin should be much longer than one millisecond.
The overall decreasing background probability is caused by the depolarization of the electron spin.
The nuclear longitudinal relaxation time $T_{1,n}$ is estimated to be $\sim100$ times of $T_{1,e}$ according to Ref.~\onlinecite{PRB_2013_Fischer}.
$T_{1,n}, T_{1,e}$ and $T_{2,n}^*$ are much longer than $T_{2,e}^*$.
Thus the off-diagonal elements representing the coherence of the electron spin decay much faster than any other elements of the density matrix.
The dephasing of the electron spin was the main effect we considered when preparing the isotropic states.

\begin{figure*}\centering
\includegraphics[width=2\columnwidth]{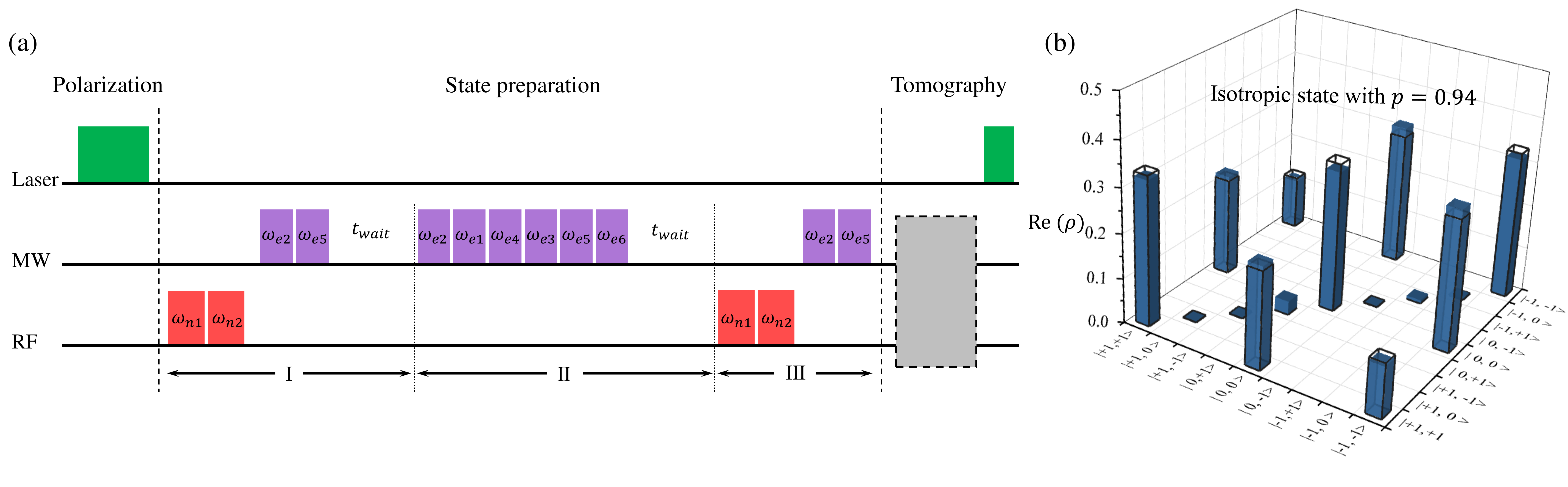}
\caption{Experimental pulse sequence and reconstructed density matrix. (a) Diagram of the pulse sequence, which includes polarization, state preparation and tomography. In the procedure of state preparation, selective MW (purple) and RF (red) pulses were performed to generate the isotropic state $\rho_{\rm iso}(p)$. Frequencies of these pulses correspond to the transition frequencies between energy levels shown in Fig.~\ref{Fig1}(b). The tomography comprises various measurement sequences to readout different elements of the density matrix (see  Supplemental Material \cite{SM} for details). The grey box stands for MW and RF pulses sequences for tomography, which is explained in detail in Supplemental Material \cite{SM}. (b) Experimental density matrix of the entangled state with $p=0.94$. The bars show the experimental outcomes while wire grids represent corresponding simulation results. Experimental results fit well with simulations and the state fidelity is $96 \%$.
 }
\label{Fig3}
\end{figure*}

Fig.~\ref{Fig3}a shows the diagram of the pulse sequence for studying the quantum correlations of the isotropic states.
It consists of three parts: polarization, state preparation and tomography.
The NV center was polarized into the state $|0,+1\rangle$ via a green laser pulse.
The preparation of the isotropic states contains three steps:
(I) applying four selective MW or RF pulses followed by a
free evolution time, $t_{wait}= 90~\mu s$, to prepare the NV center to a mixed state with the form $\rho_{\rm I}(p) = P_{+1,+1}^{\rm I}(p)|{+1,+1}\rangle\langle{+1,+1}| + P_{0,0}^{\rm I}(p) |{0,0}\rangle\langle{0,0}|+ P_{-1,-1}^{\rm I}(p)|{-1,-1}\rangle\langle{-1,-1}|$.
(II) applying six selective MW pulses and waiting the same waiting time in step (I) to manipulate the system to mixed state $\rho_{\rm II}(p) =\sum_{i,j=0,\pm1} P_{i,j}^{\rm II}(p)|i,j\rangle\langle i,j|$.
(III) applying selective MW and RF pulses to generate the off-diagonal terms of the isotropic states such that $\rho_{\rm \uppercase\expandafter{\romannumeral3}}(p) = \rho_{\rm iso}(p)$.
These selective pulses correspond to transitions between different energy levels displayed in Fig.~\ref{Fig1}b.
The quantum state tomography was performed after the state preparation and all the non-zero elements of the isotropic states were measured.
A maximum likelihood estimation method was utilized to reconstruct the density matrix of the final states \cite{PRA_2001_Daniel}.
In Fig.~\ref{Fig3}b, we show the result of the isotropic state with $p = 0.94$ as an example.
The experimental results (colored bars) fit well with the simulation results (wire grids, with the dephasing noise of the electron spin considered).
More details about the state preparation, the measurement sequences and the state reconstruction can be found in Supplemental Material \cite{SM}.

\begin{figure}\centering
\includegraphics[width=0.9\columnwidth]{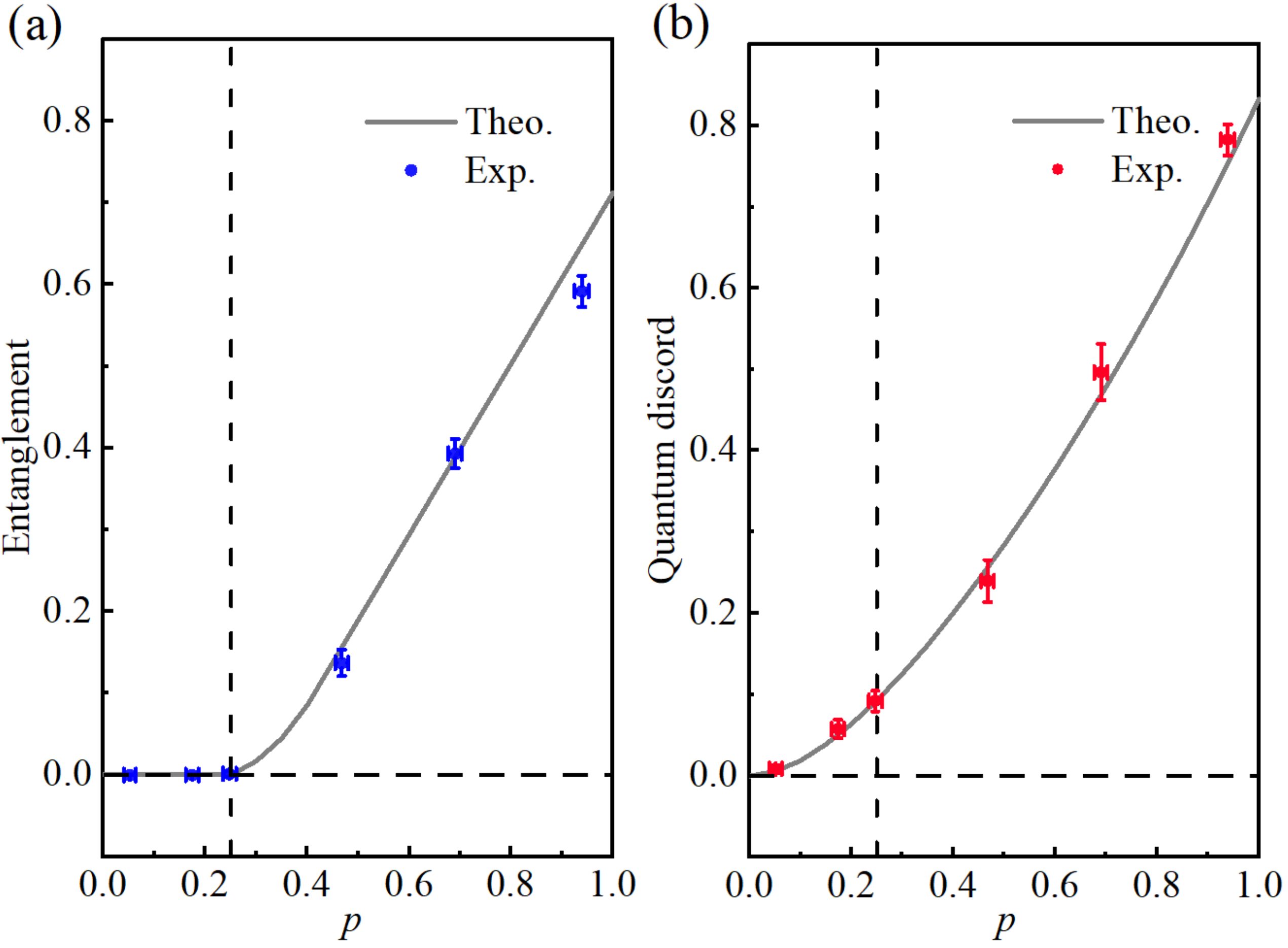}
\caption{Quantum correlations of the isotropic states.
(a) and (b) displays the experimental results of quantum entanglement and quantum discord, respectively.
Points are experimental data while solid lines show the simulation results.
The vertical error bars of the data points were calculated by Monte Carlo simulation with the Gaussian statistics.
The abscissas of the data points and their error bars were obtained by comparing the experimental results with simulations to find the most likely $p$ for each Monte Carlo run.
Some error bars are smaller than the size of dots so they are not visible.
The dashed line corresponds to $p = 1/4$, which is the dividing line between separable state and entangled state.}
\label{Fig4}
\end{figure}

The results of the quantum entanglement and quantum discord measured in our experiment are displayed in Fig.~\ref{Fig4}.
The values of the quantum discord defined by Eq.~(\ref{eq.3}) were obtained from the experimentally reconstructed density matrices using the approach introduced in Ref.~\onlinecite{PRA_2012_Rossignoli}.
The extremization in Eq.~(\ref{eq.2}) should be taken over all possible complete sets of projective measurements of subsystem $B$.
We obtained the value of the quantum discord with different measurement basis until it converged.
For the isotropic states, the measure of their entanglement can be given by the negativity defined as $N(\rho_{\rm iso}) = ({||} \rho_{\rm iso}^{PT}{||}_1 -1)/2$ where $PT$ and ${||}\cdot{||}_1$ represent partial transposition and trace norm calculation of the density matrix, respectively \cite{PRA_2007_Derkacz}.
Details about the calculation of the quantum discord and the entanglement are given in Supplemental Material \cite{SM}.
Theoretical results in Fig.~\ref{Fig4} were obtained by calculating the quantum entanglement and the quantum discord of the simulated states.
The experimental results agree well with the theoretical predictions.
In the region $0<p\le1/4$, the entanglement is zero while the non-zero quantum discord still exists.
This unambiguously means a type of quantum correlation beyond quantum entanglement had been experimentally observed in a two-qutrit system.
When $p>1/4$, the state has non-zero quantum entanglement and quantum discord that both monotonously increase with $p$.
Our results show that the qualitative behaviors of the quantum discord and the entanglement in a two-qutrit system are similar to that of a two-qubit system.
However, a smaller threshold of the quantum entanglement is observed as predicted by theories \cite{PRA_2012_Chitambar,PRA_2021_Poxleitner}.

$\mathit{Discussion}$.
Quantum correlation is an essential issue in quantum physics and recently high-dimensional quantum correlations have aroused tremendous research interest.
In this paper, we experimentally studied the quantum correlations between two qutrits in a single NV center. Entanglement in the two-qutrit system has been observed at room temperature, without resorting to cryogenic conditions \cite{arxiv_2021_Ding}. In particular, the non-classical correlations beyond entanglement in the qutrit case are founded.
These results show that NV centers are a powerful platform for further investigating the fundamental properties of high-dimensional quantum correlations, such as the essence of quantum correlations \cite{PRL_1998_Knill, TCS_2004_Biham,review_2018_De}  and their relation with quantum superposition and nonlocality \cite{review_2012_Modi,review_2018_Bera}.
Experimentally studying the dynamic behavior of high-dimensional quantum correlations and observing whether there exists sudden death \cite{science_2007_Almeida} or sudden transition \cite{PRL_2011_Auccaise} phenomena will also be interesting.
Besides, qutrit systems possess advantages in QIP \cite{review_2020_Erhard} and experimental investigations of key procedures of QIP \cite{PRL_2021_Yurtalan,PRX_2021,PRL_2021_Morvan} are necessary.
The NV center is a natural high-dimensional system and has long coherence time even at room temperature. However, it has been utilized as qubits in most studies, which indeed limited its potential.
Our work promotes the NV center as a high-dimensional system to execute quantum computation and quantum sensing tasks \cite{AQT_2019_Cozzolino, FP_2020_Wang, PRL_2021_Designolle}.
NV qutrit systems may play an important role in high-dimensional quantum information processing since further coupling two NV centers as qutrits on top of qubits \cite{science_2021,Nphys_2013,NC_2014,nature_2022} is foreseeable, albeit challenging.

This work was supported by the Chinese Academy of Sciences (Grant No. XDC07000000, No. GJJSTD20200001, No. QYZDY-SSW-SLH004, No. QYZDB-SSW-SLH005), Innovation Program for Quantum Science and Technology (Grant No. 2021ZD0302200), the National Key R$\&$D Program of China (Grant No. 2018YFA0306600), the National Natural Science Foundation of China (Grant No. 81788101, No. 11734015), Anhui Initiative in Quantum Information Technologies (Grant No. AHY050000), Hefei Comprehensive National Science Center, the Fundamental Research Funds for the Central Universities, and the open funding program from State Key Laboratory of Precision Spectroscopy (East China Normal University).
X. R thanks the Youth Innovation Promotion Association of Chinese Academy of Sciences for the support.

\onecolumngrid
\vspace{1.5cm}
\begin{center}
\newpage
\textbf{\large Supplementary Material}
\end{center}

\setcounter{figure}{0}
\setcounter{equation}{0}
\setcounter{table}{0}
\makeatletter
\renewcommand{\thefigure}{S\arabic{figure}}
\renewcommand{\theequation}{S\arabic{equation}}
\renewcommand{\thetable}{S\arabic{table}}
\renewcommand{\bibnumfmt}[1]{[RefS#1]}
\renewcommand{\citenumfont}[1]{RefS#1}

\renewcommand{\figurename}{Supplementary Figure}
\renewcommand{\tablename}{Supplementary Table}

\section{Preparation, measurement and maximum likelihood estimation of the isotropic state.}

\subsection{Preparation of the isotropic state.}

\begin{figure}[h]
\centering
\includegraphics[width=1\columnwidth]{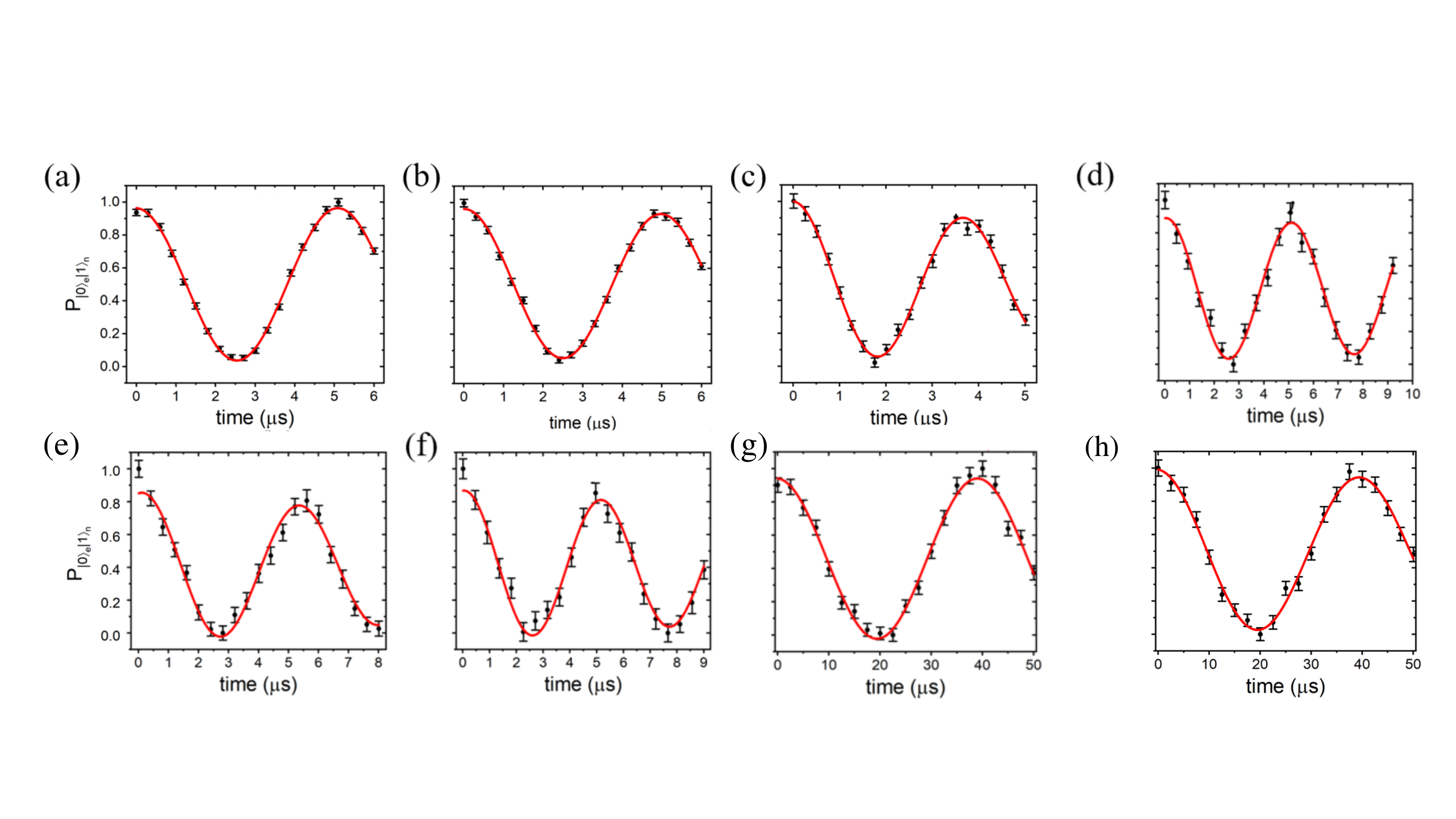}
\caption{\textbf{Rabi oscillation results.}
(a-f): Rabi oscillation results of the electron spin corresponding to transition between energy levels (a) $|{+1,+1}\rangle\leftrightarrow|{0,+1}\rangle$, (b) $|{0,+1}\rangle\leftrightarrow|{-1,+1}\rangle$, (c) $|{+1,0}\rangle\leftrightarrow|{0,0}\rangle$, (d) $|{0,0}\rangle\leftrightarrow|{-1,0}\rangle$,  (e) $|{+1,-1}\rangle\leftrightarrow|{0,-1}\rangle$ and  (f) $|{0,-1}\rangle\leftrightarrow|{-1,-1}\rangle$, respectively.
All the Rabi frequencies were calibrated to ${\rm \Omega_{MW}}=0.2$ MHz.
(g-h): Rabi oscillation results of the nuclear spin corresponding to transition between energy levels (g) $|{0,+1}\rangle\leftrightarrow|{0,0}\rangle$  and  (h) $|{0,0}\rangle\leftrightarrow|{0,-1}\rangle$.
These Rabi frequencies were calibrated to ${\rm \Omega_{RF}}=25$ kHz.
}
\label{Rabi oscillations}
\end{figure}

The two-qutrit isotropic state $\rho_{\rm iso}$ depending on the parameter $p$ is given by

\begin{equation}
\rho_{\rm iso}(p) = \frac{(1-p)}{9} \mathbb{I}_3 \otimes \mathbb{I}_3 +p |\psi\rangle \langle \psi| \ \ ,
\end{equation}
where $p\in[0,1]$ and $|\psi\rangle =  (|{+1,+1}\rangle +|{0,0}\rangle+|{-1,-1}\rangle)/\sqrt{3}$ is the maximally entangled state.
In the two-qutrit system based on the NV center, the energy levels $|+1\rangle_e$, $|0\rangle_e$ and $|-1\rangle_e$ ($|+1\rangle_n$, $|0\rangle_n$ and $|-1\rangle_n$) correspond to the state $|+1\rangle, |0\rangle$ and $|-1\rangle$ of the first (second) qutrit, respectively.
In this representation, the density matrix of $\rho_{\rm iso}$ takes the form
\begin{equation}
\rho_{\rm iso}(p) =
\begin{pmatrix}
\frac{1+2p}{9} & 0 & 0 & 0 & \frac{p}{3} & 0 & 0 & 0 & \frac{p}{3}\\
              0 &  \frac{1-p}{9} & 0 & 0 & 0 & 0 & 0 & 0 & 0\\
              0 & 0 &  \frac{1-p}{9} & 0 & 0 & 0 & 0 & 0 & 0\\
              0 & 0 & 0 &  \frac{1-p}{9} & 0 & 0 & 0 & 0 & 0\\
              \frac{p}{3} & 0 & 0 & 0 & \frac{1+2p}{9} & 0 & 0 & 0 & \frac{p}{3}\\
              0 & 0 & 0 & 0 & 0 &  \frac{1-p}{9} & 0 & 0 & 0\\
              0 & 0 & 0 & 0 & 0 & 0 &  \frac{1-p}{9} & 0 & 0\\
              0 & 0 & 0 & 0 & 0 & 0 & 0 &  \frac{1-p}{9} & 0\\
              \frac{p}{3} & 0 & 0 & 0 & \frac{p}{3} & 0 & 0 & 0 & \frac{1+2p}{9}
\end{pmatrix} \ .
\end{equation}

MW and RF pulses were applied to manipulate the state of the two-qutrit system.
The Rabi frequencies of the MW and RF pulses used in our experiment were calibrated to ${\rm \Omega_{MW}}=0.2$ MHz and ${\rm \Omega_{RF}}=25$ kHz, respectively.
The experimental results of these Rabi oscillations are displayed in Fig.~\ref{Rabi oscillations}.

The NV center was polarized into the state $|0,+1\rangle$ via laser pulses at the beginning of the state preparation.
Then the following three steps were performed to prepare the $\rho_{\rm iso}$. Information of the pulses used in the preparation process is displayed in Table.~\ref{state preparation parameters}.
\begin{itemize}
\item[(i)]
Two selective RF pulses with frequency $\omega_{n1}$ and $\omega_{n2}$ respectively were applied in sequence to manipulate the NV center into the state $|\psi_1\rangle = \cos\frac{\alpha_1}{2}|{0,+1}\rangle +\sin\frac{\alpha_1}{2}\cos\frac{\alpha_2}{2}|0,0\rangle+\sin\frac{\alpha_1}{2}\sin\frac{\alpha_2}{2}|{0,-1}\rangle$.
Then two selective MW pulses with frequency $\omega_{e2}$ and $\omega_{e5}$ respectively were performed to prepare the NV center into the state $|\psi_2\rangle = \cos\frac{\alpha_1}{2}|{+1,+1}\rangle +\sin\frac{\alpha_1}{2}\cos\frac{\alpha_2}{2}|0,0\rangle+\sin\frac{\alpha_1}{2}\sin\frac{\alpha_2}{2}|{-1,-1}\rangle$. The values of $\alpha_{j}$ ($j = 1,2$) are displayed in the Table.~\ref{state preparation parameters}.
After a waiting time of $t_{wait} = 90~\mu$s, the coherence of the electron spin dissipated, resulting in the NV center in the mixed state with the form
\begin{equation}
\rho_{\rm I}(p) =
\begin{pmatrix}
\frac{1-p}{3} & 0 & 0 & 0 & 0 & 0 & 0 & 0 & 0\\
0 & 0 & 0 & 0 & 0 & 0 & 0 & 0 & 0\\
0 & 0 & 0 & 0 & 0 & 0 & 0 & 0 & 0\\
0 & 0 & 0 & 0 & 0 & 0 & 0 & 0 & 0\\
0 & 0 & 0 & 0 & \frac{1+2p}{3} & 0 & 0 & 0 & 0\\
0 & 0 & 0 & 0 & 0 & 0 & 0 & 0 & 0\\
0 & 0 & 0 & 0 & 0 & 0 & 0 & 0 & 0\\
0 & 0 & 0 & 0 & 0 & 0 & 0 & 0 & 0\\
0 & 0 & 0 & 0 & 0 & 0 & 0 & 0 & \frac{1-p}{3}
\end{pmatrix} \ .
\end{equation}
\item[(ii)]
 Six selective MW pulses with pulse phase being X  were applied in sequence as shown in Fig.3(a) of main text. The rotation angles of these pulses $\theta_{j}$ ($j = 1,2,3,4,5,6$) are shown in the Table.~\ref{state preparation parameters}.
 Then after the same waiting time $t_{wait} = 90\ \mu s$, $\rho_{\rm I}(p)$  evolved to a mixed state with the form
\begin{equation}
\rho_{\rm II}(p) =
\begin{pmatrix}
\frac{1-p}{9} & 0 & 0 & 0 & 0 & 0 & 0 & 0 & 0\\
0 & \frac{1-p}{9} & 0 & 0 & 0 & 0 & 0 & 0 & 0\\
0 & 0 & \frac{1-p}{9} & 0 & 0 & 0 & 0 & 0 & 0\\
0 & 0 & 0 & \frac{1-p}{9} & 0 & 0 & 0 & 0 & 0\\
0 & 0 & 0 & 0 & \frac{1+8p}{9} & 0 & 0 & 0 & 0\\
0 & 0 & 0 & 0 & 0 & \frac{1-p}{9} & 0 & 0 & 0\\
0 & 0 & 0 & 0 & 0 & 0 & \frac{1-p}{9} & 0 & 0\\
0 & 0 & 0 & 0 & 0 & 0 & 0 & \frac{1-p}{9} & 0\\
0 & 0 & 0 & 0 & 0 & 0 & 0 & 0 & \frac{1-p}{9}
\end{pmatrix} \ .
\end{equation}

\item[(iii)]
Two selective RF pulses and two selective MW pulses were applied in sequence to generate the off-diagonal terms of the isotropic state.
The rotation angles of these pulses are shown in the Table.~\ref{state preparation parameters}.
Then the state of the NV center was $\rho_{\rm III}(p) = \rho_{\rm iso}(p)$.
So the isotropic state was generated in the two-qutrit system constructed by the NV center.

\end{itemize}

\begin{table}

\caption{\textbf{ Detailed information about the parameters of the MW and RF pulses used in the process of the isotropic state preparation.}}

\begin{tabular}{p{1.5cm}<{\centering}|c|p{4cm}<{\centering}|p{5cm}<{\centering}}

\toprule

step & pulse frequency & transition states & pulse phase and rotation angle \\

\midrule

\multirow{4}{* }{$\rm I$} &  $ \omega_{n,1}$ & $|0,+1\rangle\leftrightarrow |0,0\rangle$      & X ,  $\alpha_1=2\arcsin \sqrt{\frac{2+p}{3}}$ \\

&$   \omega_{n,2}$ & $|0,0\rangle\leftrightarrow |0,-1\rangle$      & X ,  $\alpha_2=2\arcsin \sqrt{\frac{1-p}{2+p}}$ \\

&$   \omega_{e,2}$ & $|0,+1\rangle\leftrightarrow |+1,+1\rangle$    & X ,  $\pi$ \\

&$   \omega_{e,5}$ & $|0,-1\rangle\leftrightarrow |-1,-1\rangle$    & X ,  $\pi$ \\

\midrule

\multirow{6}{*}{$\rm II$} & $ \omega_{e,2}$ & $|0,+1\rangle\leftrightarrow |+1,+1\rangle$    & X ,  $\theta_1=2\arcsin \sqrt{\frac{2}{3}}$ \\

 & $  \omega_{e,1}$ & $|0,+1\rangle\leftrightarrow |-1,+1\rangle$    & X ,  $\theta_2=\pi/2$ \\

&$  \omega_{e,4}$ & $|0,0\rangle\leftrightarrow |+1,0\rangle$      & X ,  $\theta_3=2\arcsin \sqrt{\frac{1-p}{3+6p}}$ \\

&$  \omega_{e,3}$ & $|0,0\rangle\leftrightarrow |-1,0\rangle$      & X ,  $\theta_4=2\arcsin \sqrt{\frac{1-p}{2+7p}}$ \\

&$  \omega_{e,5}$ & $|0,-1\rangle\leftrightarrow |-1,-1\rangle$    & X ,  $\theta_5=2\arcsin \sqrt{\frac{2}{3}}$ \\

&$  \omega_{e,6}$ & $|0,-1\rangle\leftrightarrow |+1,-1\rangle$    & X ,  $\theta_6=\pi/2$ \\

\midrule

\multirow{4}{*}{$\rm III$}&$\omega_{n,1}$ & $|0,+1\rangle\leftrightarrow |0,0\rangle$      & Y ,  $\phi_1=2\arcsin \sqrt{\frac{1}{3}}$ \\

&$ \omega_{n,2}$ & $|0,0\rangle\leftrightarrow |0,-1\rangle$      & Y ,  $\phi_2=\pi/2$ \\

&$ \omega_{e,2}$ & $|0,+1\rangle\leftrightarrow |+1,+1\rangle$    & Y ,  $\pi$ \\

&$\omega_{e,5}$ & $|0,-1\rangle\leftrightarrow |-1,-1\rangle$    & Y ,  $\pi$ \\

\bottomrule

\end{tabular}

\label{state preparation parameters}

\end{table}

\subsection{Measurement of the isotropic state.}

The isotropic state was measured via a series of measurement sequences.
The photoluminescence (PL) rate of the NV center depends on the state of both the electron spin and the nuclear spin after being pumped by the laser pulse \cite{PRL_V.Jacques}.
This property was utilized to readout the state of the two-qutrit system.
For simplicity, eigenstates $|+1,+1\rangle$, $|+1,0\rangle$, $|+1,-1\rangle$, $|0,+1\rangle$, $|0,0\rangle$, $|0,-1\rangle$, $|-1,+1\rangle$, $|-1,0\rangle$ and $|-1,-1\rangle$ are relabeled as $|1\rangle$, $|2\rangle$, $|3\rangle$, $|4\rangle$, $|5\rangle$, $|6\rangle$, $|7\rangle$, $|8\rangle$ and $|9\rangle$, respectively in the following.

\begin{figure}[h]
\centering
\includegraphics[width=0.8\columnwidth]{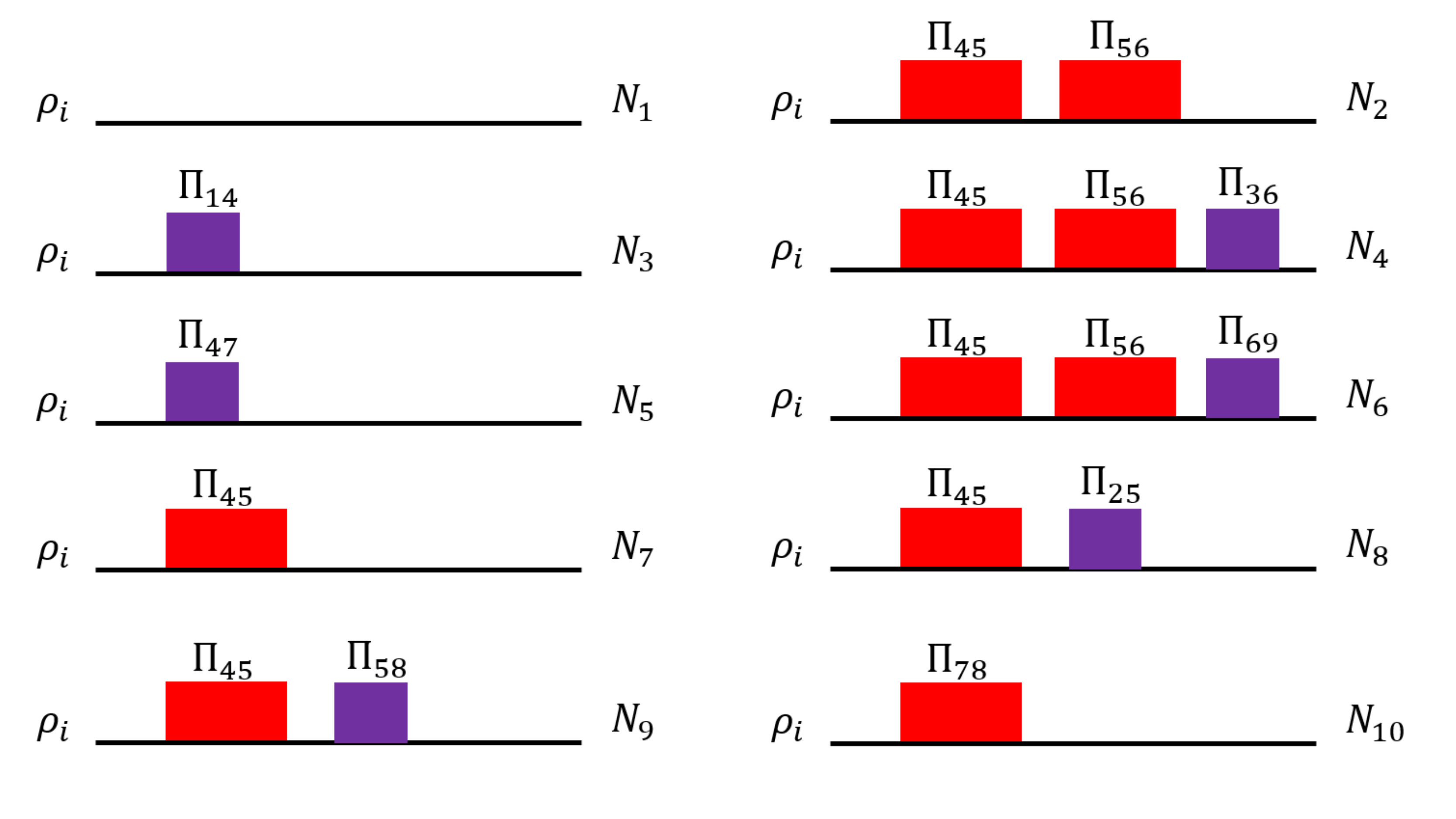}
\caption{\textbf{Normalization pulse sequences.}
$\rho_i$ denotes the initialized state after a optical excitation.
${\rm \Pi}_{m,n}$ denotes the selective $\pi$ pulse applied between state $|m\rangle$ and state $|n\rangle$.
$N_j$ ($j\in[1,10]$) indicates the detected PL rate  after applying these pulse sequence.
}
\label{normalization sequences}
\end{figure}

We measured the PL rates of different eigenstates via pulse sequences given in Fig.~\ref{normalization sequences}.
Firstly, after performing a laser pulse, the NV center was prepared into the state
 $\rho_i = \lambda|1\rangle\langle1|+p_e|4\rangle\langle4|+\lambda|7\rangle\langle7|$ with $p_e$ denoting the polarization of the electron spin and $\lambda = (1-p_e)/2$.
Polarization of the nuclear spin was high and can be  regarded as 1 here.
When all the pulses sequences shown in Fig.~\ref{normalization sequences} are taken into account,
we can obtain the following equations
\begin{equation}
\label{normalization_equations}
\begin{bmatrix}
\lambda & 0 & 0 & p_e  &  0  &  0  &  \lambda & 0 & 0   \\
\lambda & 0 & 0 &  0   &  0  & p_e &  \lambda & 0 & 0   \\
  p_e & 0 & 0 & \lambda  &  0  &  0  &  \lambda & 0 & 0 \\
\lambda & 0 & p_e &  0  &  0  &  0  &  \lambda & 0 & 0   \\
\lambda & 0 & 0 &  \lambda &  0  &  0  &  p_e & 0 & 0   \\
\lambda & 0 & 0 & 0  &  0  &  0  &  \lambda & 0 & p_e   \\
\lambda & 0 & 0 & 0  &  p_e  &  0  &  \lambda & 0 & 0   \\
\lambda & p_e & 0 & 0  &  0  &  0  &  \lambda & 0 & 0   \\
\lambda & 0 & 0 & 0  &  0  &  0  &  \lambda & p_e & 0   \\
\lambda & 0 & 0 & p_e  &  0  &  0  &  0 & \lambda & 0   \\
\end{bmatrix}
\begin{bmatrix}
L_{|1\rangle} \\ L_{|2\rangle} \\ L_{|3\rangle} \\ L_{|4\rangle} \\ L_{|5\rangle} \\ L_{|6\rangle} \\ L_{|7\rangle} \\ L_{|8\rangle} \\ L_{|9\rangle}
\end{bmatrix}
= \begin{bmatrix}
N_1 \\ N_2 \\ N_3 \\ N_4 \\N_5 \\N_6 \\ N_7 \\ N_8 \\ N_9 \\ N_{10}
\end{bmatrix} \ ,
\end{equation}
where $L_{|m\rangle}$ denotes the PL rate of state $|m\rangle$ ($m\in [1,9]$), $N_j$ is the PL rate of the state obtained via applying the pulse sequence in Fig.~\ref{normalization sequences} to initialized state $\rho_i$.
The polarization $p_e$ and the PL rates $L_{|m\rangle}$ can be obtained by solving equation (\ref{normalization_equations}).

Suppose that the density matrix of the isotropic state we prepared takes the form
\begin{equation}
\rho_f =
\begin{pmatrix}
\rho_{11} & \rho_{12} & \rho_{13} & \rho_{14} & \rho_{15} & \rho_{16} & \rho_{17} & \rho_{18} & \rho_{19}\\
\rho_{12}^* &  \rho_{22} & \rho_{23} & \rho_{24} & \rho_{25} & \rho_{26} & \rho_{27} & \rho_{28} & \rho_{29}\\
\rho_{13}^* & \rho_{23}^* &  \rho_{33} & \rho_{34} & \rho_{35} & \rho_{36} & \rho_{37} & \rho_{38} & \rho_{39}\\
\rho_{14}^* & \rho_{24}^* & \rho_{34}^* &  \rho_{44} & \rho_{45} & \rho_{46} & \rho_{47} & \rho_{48} & \rho_{49}\\
\rho_{15}^* & \rho_{25}^* & \rho_{35}^* & \rho_{45}^* & \rho_{55} & \rho_{56} & \rho_{57} & \rho_{58} & \rho_{59}\\
\rho_{16}^* & \rho_{26}^* & \rho_{36}^* & \rho_{46}^* & \rho_{56}^* &  \rho_{66} & \rho_{67} & \rho_{68} & \rho_{69}\\
\rho_{17}^* & \rho_{27}^* & \rho_{37}^* & \rho_{47}^* & \rho_{57}^* & \rho_{67}^* &  \rho_{77} & \rho_{78} & \rho_{79}\\
\rho_{18}^* & \rho_{28}^* & \rho_{38}^* & \rho_{48}^* & \rho_{58}^* & \rho_{68}^* & \rho_{78}^* &  \rho_{88} & \rho_{89}\\
\rho_{19}^* & \rho_{29}^* & \rho_{39}^* & \rho_{49}^* & \rho_{59}^* & \rho_{69}^* & \rho_{79}^* & \rho_{89}^* & \rho_{99}
\end{pmatrix} \ ,
\end{equation}
and some off-diagonal elements of the density matrix were redefined as
\begin{equation}
\left\{
\begin{aligned}
\rho_{15} &= \mu_1+i\nu_1  \\
\rho_{59} &= \mu_2+i\nu_2  \\
\rho_{19} &= \mu_3+i\nu_3  \\
\end{aligned}
\right. \ .
\end{equation}

\begin{figure}
\centering
\includegraphics[width=0.8\columnwidth]{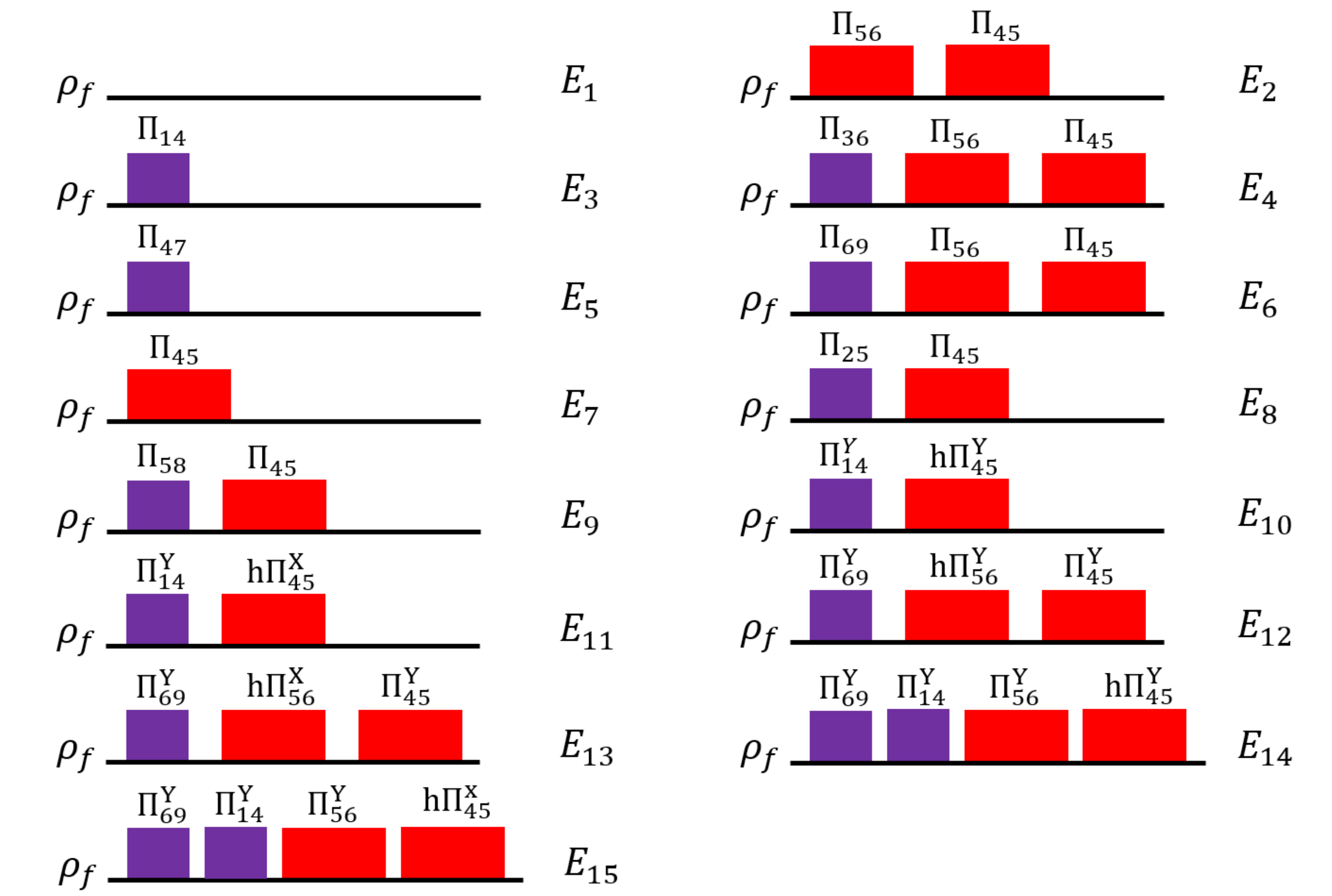}
\caption{\textbf{Measurement pulse sequences of the isotropic state.}
$\rho_f$ denotes the density matrix of the prepared isotropic state.
${\rm \Pi}_{m,n}^{\rm X(Y)}$ (${\rm h}{\rm \Pi}_{m,n}^{\rm X(Y)}$) denotes the selective $\pi$ ($\pi/2$) pulse applied between state $|m\rangle$ and state $|n\rangle$ along X (Y) axis.
$E_j$ ($j\in[1,15]$) indicates the detected PL rate  after applying these pulse sequence.
}
\label{tomography sequences}
\end{figure}

From the measurement sequences displayed in Fig.~\ref{tomography sequences}, we can obtain the following equations
\begin{equation}
\label{measurement_equations}
\begin{aligned}
\addtocounter{MaxMatrixCols}{10}
&
\begin{bmatrix}
L_1    & L_2 & L_3 & L_4  &  L_5    &  L_6  &  L_7 & L_8 & L_9    & 0     & 0       & 0      & 0      & 0       & 0  \\
L_1    & L_2 & L_3 & L_5  &  L_6    &  L_4  &  L_7 & L_8 & L_9    & 0     & 0       & 0      & 0      & 0       & 0  \\
L_4    & L_2 & L_3 & L_1  &  L_5    &  L_6  &  L_7 & L_8 & L_9    & 0     & 0       & 0      & 0      & 0       & 0  \\
L_1    & L_2 & L_4 & L_5  &  L_6    &  L_3  &  L_7 & L_8 & L_9    & 0     & 0       & 0      & 0      & 0       & 0  \\
L_1    & L_2 & L_3 & L_7  &  L_5    &  L_6  &  L_4 & L_8 & L_9    & 0     & 0       & 0      & 0      & 0       & 0  \\
L_1    & L_2 & L_3 & L_5  &  L_6    &  L_9  &  L_7 & L_8 & L_4    & 0     & 0       & 0      & 0      & 0       & 0  \\
L_1    & L_2 & L_3 & L_5  &  L_4    &  L_6  &  L_7 & L_8 & L_9    & 0     & 0       & 0      & 0      & 0       & 0  \\
L_1    & L_4 & L_3 & L_5  &  L_2    &  L_6  &  L_7 & L_8 & L_9    & 0     & 0       & 0      & 0      & 0       & 0  \\
L_1    & L_2 & L_3 & L_5  &  L_8    &  L_6  &  L_7 & L_4 & L_9    & 0     & 0       & 0      & 0      & 0       & 0  \\
L_{11} & L_2 & L_3 & L_1  & L_{11}  &  L_6  &  L_7 & L_8 & L_9    &L_{13} & 0       & 0      & 0      & 0       & 0  \\
L_{11} & L_2 & L_3 & L_1  & L_{11}  &  L_6  &  L_7 & L_8 & L_9    & 0     & L_{13}  & 0      & 0      & 0       & 0  \\
L_1    & L_2 & L_3 & L_5  & L_{11}  &  L_9  &  L_7 & L_8 & L_{12} & 0     & 0       & L_{14} & 0      & 0       & 0  \\
L_1    & L_2 & L_3 & L_5  & L_{11}  &  L_9  &  L_7 & L_8 & L_{12} & 0     & 0       & 0      & L_{14} & 0       & 0  \\
L_{11} & L_2 & L_3 & L_1  &  L_6    &  L_9  &  L_7 & L_8 & L_{11} & 0     & 0       & 0      & 0      & L_{13}  & 0  \\
L_{11} & L_2 & L_3 & L_1  &  L_6    &  L_9  &  L_7 & L_8 & L_{11} & 0     & 0       & 0      & 0      & 0       & L_{13}
\end{bmatrix}
\cdot
\begin{bmatrix}
\rho_{11} \\ \rho_{22} \\ \rho_{33} \\ \rho_{44} \\ \rho_{55} \\ \rho_{66} \\ \rho_{77} \\ \rho_{88} \\ \rho_{99} \\ \mu_1 \\ \nu_1 \\ \mu_2 \\ \nu_2 \\ \mu_3 \\ \nu_3
\end{bmatrix}
= \begin{bmatrix}
E_1 \\ E_2 \\ E_3 \\ E_4 \\E_5 \\E_6 \\ E_7 \\ E_8 \\ E_9 \\ E_{10}  \\ E_{11} \\ E_{12} \\ E_{13} \\ E_{14} \\ E_{15}
\end{bmatrix}\ ,
\end{aligned}
\end{equation}
where
\begin{equation}
\left\{
\begin{aligned}
L_{11} & = (L_4+L_5)/2 \\
L_{12} & = (L_4+L_6)/2 \\
L_{13} & = L_5 - L_4   \\
L_{14} & = L_4 - L_6
\end{aligned}
\right. \ ,
\end{equation}
and $E_j$ is the PL rate of the state obtained via applying the pulse sequence in Fig.~\ref{tomography sequences} to state $\rho_f$.
The theoretical non-zero elements of the isotropic state can be obtained by solving equation (\ref{measurement_equations}).
We also measured a few matrix elements that were supposed to be zero theoretically and the experimental data show that these elements were very close to zero.

\subsection{Maximum likelihood estimation of the isotropic state.}

The obtained density matrix can violate basic physical properties, so we employed a maximum likelihood estimation (MLE) method to acquire the most possible physical state \cite{PRA_2001_Daniel}.
The likelihood function we adopted reads:
\begin{align}
       \mathcal{L} = \sum_{ij} \frac{(\sigma_{ij}-\rho_{ij})^2}{2\sigma_{ij}},
       \label{l-function}
\end{align}
where $\rho_{ij}$ ($\sigma_{ij}$) is the element of the experimental (estimated) density matrix.
The matrix $\sigma$ is constructed by
\begin{align}
\sigma &= \frac{T^{\dagger}(k)T(k)}{{\rm tr}(T^{\dagger}(k)T(k))} ,
\end{align}
with
\begin{align}
T(k) & =  \left(\begin{array}{ccccccccc}
k_1    & 0  & 0 & 0 & 0 & 0 & 0 & 0 & 0  \\
   0   &k_2 & 0 & 0 & 0 & 0 & 0 & 0 & 0  \\
   0   & 0 &k_3 & 0 & 0 & 0 & 0 & 0 & 0  \\
   0   & 0 & 0 &k_4 & 0 & 0 & 0 & 0 & 0  \\
k_{10} & 0 & 0 & 0 & k_5& 0 & 0 & 0 & 0  \\
   0   & 0 & 0 & 0 & 0 &k_6 & 0 & 0 & 0 \\
   0   & 0 & 0 & 0 & 0 & 0 &k_7 & 0 & 0 \\
   0   & 0 & 0 & 0 & 0 & 0 & 0 & k_8& 0  \\
k_{11} & 0 & 0 & 0 &k_{12} & 0 & 0 & 0 & k_9
\end{array}
\right) \notag,
\end{align}
where $k_i$ ($i\in\{1,2,3,...,12\}$) is independent real parameter.
The final physical state can be reconstructed by varying parameter $k_i$ to minimize the value of the Eq. (\ref{l-function}).

The photon counts of the measurement sequences fluctuate due to the shot noise, which leads to uncertainties of the elements of the experimental reconstructed density matrix.
We generated 100 different experimental density matrices ($\rho^M$, $M\in\{1,2,3,\dots,100\}$) by Monte Carlo Simulations with Gaussian statistics.
Maximum likelihood estimation was applied for each $\rho^M$ to obtain the corresponding $\sigma^M$.
The average estimated density matrix $\overline{\sigma^M}$ was calculated as the final state (Fig.3b in the main text as an instance).
We calculated the fidelity $F(\sigma^M,\rho_{\rm simu}(p))$  to find $p^M$ that maximizes $F(\sigma^M,\rho_{\rm simu}(p))$.
Then the horizontal coordinates of the data in Fig.4 of the main text equal the average value of $p^M$ and the horizontal error bars are the standard derivation of $p^M$.

\section{Calculation of the quantum discord and the quantum entanglement}

The QD is defined by Eq.(3) in the main text.
It is challenging to calculate the QD of a $d$-level system because the optimization needs to be executed in the parameter space with dimension $d(d-1)$ \cite{review_2018_Bera}.
We used the method introduced in Ref.~\onlinecite{PRA_2012_Rossignoli} to calculate the QD of our two-qutrit system, where the measurement operators were parameterized by six parameters $\alpha$, $\beta$, $\gamma$, $\psi$, $\theta$ and $\phi$.
Parameters $\alpha,\beta\in [0,\pi]$ and $\gamma \in (-\pi/2,\pi/2]$ were used to construct a group of orthonormal bases of a qutrit system given by
\begin{equation}
\left\{
\begin{aligned}
|{+1}_r\rangle &= \cos \beta(e^{-i\phi_0}\cos\alpha|{+1}\rangle + e^{i\phi_0}\sin\alpha|{-1}\rangle)-\sin\beta e^{-i\gamma}|0\rangle \notag \\
|0_r\rangle & = \sin\beta(e^{-i\phi_0}\cos\alpha|{+1}\rangle +e^{i\phi_0}\sin\alpha|{-1}\rangle) + \cos \beta e^{-i\gamma}|0\rangle \notag \\
|{-1}_r\rangle & = -e^{-i\phi_0}\sin\alpha|{+1}\rangle+e^{i\phi_0}\cos\alpha|{-1}\rangle
\end{aligned}
\right. \ ,
\label{measurement bases}
\end{equation}
where $\tan\phi_0 = \tan\gamma \tan(\pi/4-\alpha)$.
Then parameters $\psi,\theta,\phi \in [0,2\pi]$ were used to define a general rotation of a qutrit system with the form
\begin{equation}
R(\psi,\theta,\phi)= e^{-i\psi S_z}e^{-i\theta S_y}e^{-i\phi S_z},
\end{equation}
where $S_x$, $S_y$ and $S_z$ are the spin operators of a spin-1 system.
A set of orthogonal measurement bases of a qutrit system can be constructed by applying rotation $R(\psi,\theta,\phi)$ to the basis given in equation (\ref{measurement bases}).

The QD was obtained by traversing all the projective measurements with the basis defined above.
We discretized the parameter space and used MATLAB to execute the calculations.
With the increment of the discretization of the parameters, the calculation converges and then the QD was obtained.

The quantum entanglement was characterized by the negativity \cite{PRA_2007_Derkacz} of the density matrix defined as
\begin{equation}
N(\rho_{\rm iso}) = \frac{{||} \rho_{\rm iso}^{PT}{||}_1 -1}{2},
\end{equation}
where $\rho_{\rm iso}^{PT}$ is the partial transposition of state $\rho_{\rm iso}$ and $||\cdot||_1$ denotes the trace norm.
We calculated the quantum discord and the quantum entanglement for isotropic states with different values of the parameter $p$ and the results are shown in Fig.4 of the main text.

\section{ Alignment of the magnetic field to the NV symmetry axis and polarization of the nuclear spin.}

\begin{figure}[h]
\centering
\includegraphics[width=0.7\columnwidth]{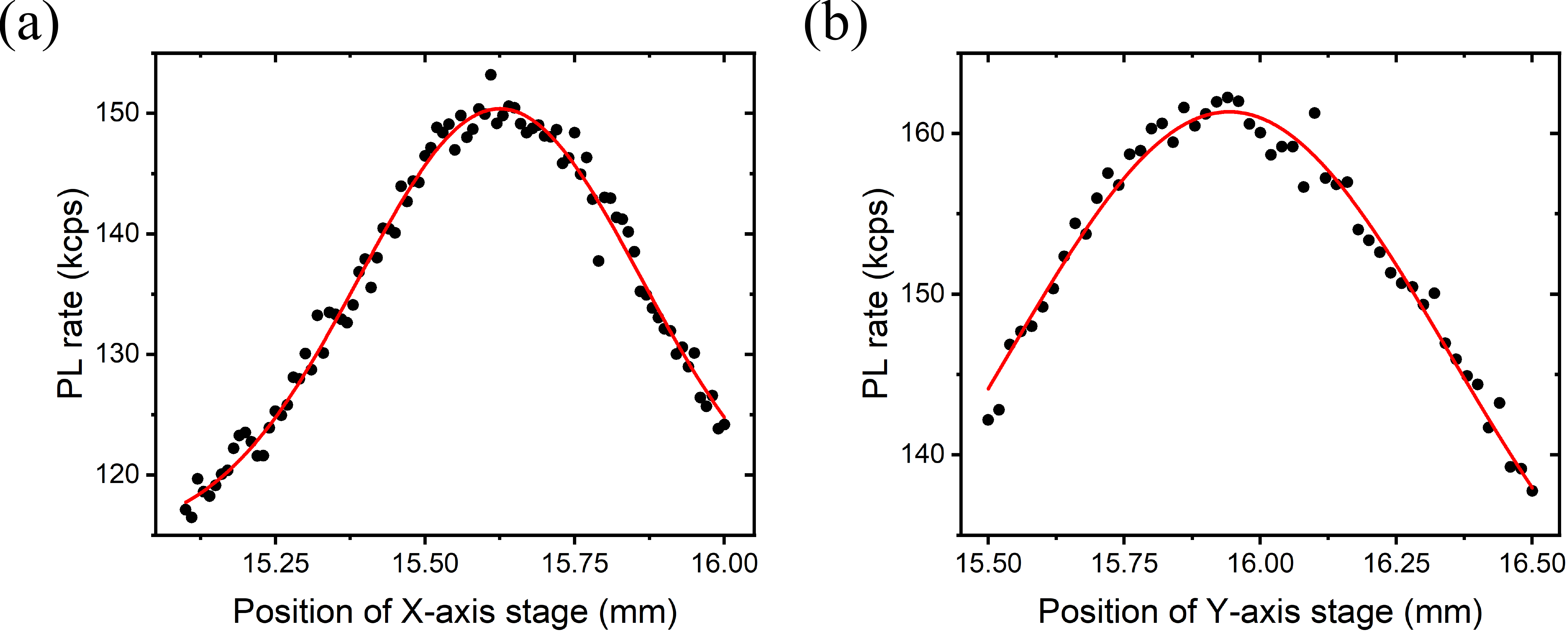}
\caption{\textbf{Alignment of the magnetic field.}
(a) and (b) shows the PL rate of the NV center versus the position of the X-axis and Y-axis stages of the magnet respectively.
Gauss fitting of the data gives the position of the magnet that the direction of the magnetic field most aligns with the NV symmetry axis.
}
\label{alignment of B field}
\end{figure}

At around 500 G, the PL rate of the NV center drops rapidly when the direction of the magnetic field deviates from the NV symmetry axis \cite{PRL_V.Jacques}.
This property was utilized to realize refined alignment of the magnetic field to the NV symmetry axis.
We adjusted the position of the magnet and meanwhile recorded the PL rate of the NV center to find the best position of the magnet.
The results are shown in Fig.\ref{alignment of B field}.

\begin{figure}[h]
\centering
\includegraphics[width=0.9\columnwidth]{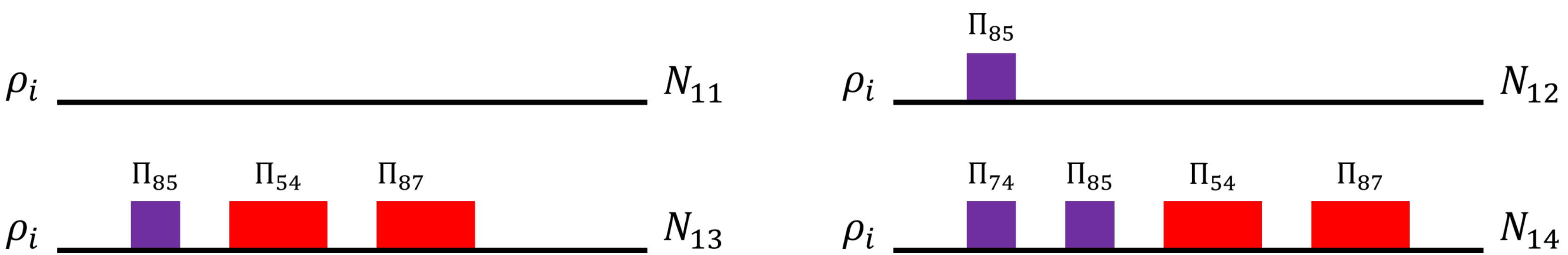}
\caption{\textbf{Measurement of the nuclear spin polarization.}
$\rho_i$ denotes the initialized state after an optical excitation.
${\rm \Pi}_{m,n}$ denotes the selective $\pi$ pulse applied between state $|m\rangle$ and state $|n\rangle$.
$N_j$ ($j\in[11,14]$) indicates the detected PL rate  after applying these pulse sequence.
}
\label{Nuclear_Polarization_Sequence}
\end{figure}

Then, following the method introduced in the supplementary material of \cite{Nature_2012_Sar}, we measured the polarization of the nuclear spin via pulse sequences shown in Fig \ref{Nuclear_Polarization_Sequence}.
The pulse sequences in Fig \ref{Nuclear_Polarization_Sequence} lead to the following equations
\begin{equation}
\left\{
\begin{aligned}
N_{11} &= \lambda\cdot p_n\cdot l_1+\lambda\cdot(1-p_n)\cdot l_2+p_e\cdot p_n\cdot l_4+p_e\cdot(1-p_n)\cdot l_5+\lambda\cdot p_n\cdot l_7+\lambda\cdot(1-p_n)\cdot l_8 \notag \\
N_{12} & = \lambda\cdot p_n\cdot l_1+\lambda\cdot(1-p_n)\cdot l_2+p_e\cdot p_n\cdot l_4+\lambda\cdot(1-p_n)\cdot l_5+\lambda\cdot p_n\cdot l_7+p_e\cdot(1-p_n)\cdot l_8 \notag \\
N_{13} & = \lambda\cdot p_n\cdot l_1+\lambda\cdot(1-p_n)\cdot l_2+\lambda\cdot(1-p_n)\cdot l_4+p_e\cdot p_n\cdot l_5+p_e\cdot(1-p_n)\cdot l_7+\lambda\cdot p_n\cdot l_8  \\
N_{14} & = \lambda\cdot p_n\cdot l_1+\lambda\cdot(1-p_n)\cdot l_2+\lambda\cdot(1-p_n)\cdot l_4+\lambda\cdot p_n\cdot l_5+p_e\cdot(1-p_n)\cdot l_7+p_e\cdot p_n\cdot l_8
\end{aligned}
\right. \ .
\label{measurement bases}
\end{equation}
Solving the equations above, we can obtain
\begin{equation}
p_n=\frac{N_{13}-N_{14}}{N_{11}-N_{12}+N_{13}-N_{14}} .
\end{equation}
Our experimental result shows that $p_n=0.981(5)$ for three million loops of these pulse sequences.
The high polarization of the nuclear spin also indicates that the misalignment of the magnetic field to the NV symmetry axis should be very small since the nuclear polarization also decreases rapidly when the direction of the magnetic field deviates from the NV symmetry axis.

When calculating the experimental density matrix of the isotropic state $\rho_{\rm iso}$ in section $\rm{\uppercase\expandafter{\romannumeral1}}$, we assumed that $p_n=1$.
Utilizing $p_n=0.981$, a modified density matrix can be obtained as $\rho_{\rm iso}^\prime(p)$.
The difference between $\rho_{\rm iso}$ and $\rho_{\rm iso}^\prime(p)$ is very subtle.
For example, when $p=0.5$, the fidelity between $\rho_{\rm iso}$ and $\rho_{\rm iso}^\prime(p)$ is 0.9996.
Thus the assumption $p_n=1$ is reasonable.


\begin{thebibliography}{99}



\bibitem{JPhys_2016_Adesso} G. Adesso, T. R. Bromley, and M. Cianciaruso, Measures and applications of quantum correlations, \emph{J. Phys. A} \textbf{49}, 473001 (2016).

\bibitem{review_2018_De} G. De Chiara and A. Sanpera, Genuine quantum correlations in quantum manybody systems: a review of recent progress, \emph{Rep. Prog. Phys.} \textbf{81}, 074002 (2018).

\bibitem{book_2015_streltsov} A. Streltsov, Quantum Correlations Beyond Entanglement and Their Role in Quantum Information Theory (Springer, New York, 2015).

\bibitem{review_2018_Braun} D. Braun, G. Adesso, F. Benatti, R. Floreanini, U. Marzolino, M. W. Mitchell, and S. Pirandola, Quantum-enhanced measurements without entanglement, \emph{Rev. Mod. Phys.} \textbf{90}, 035006 (2018).

\bibitem{review_2009} R. Horodecki, P. Horodecki, M. Horodecki, and K. Horodecki, Quantum entanglement, \emph{Rev. Mod. Phys.} \textbf{81}, 865 (2009).

\bibitem{PRL_2001_Olivier} H. Ollivier and W. H. Zurek, Quantum discord: a measure of the quantumness of correlations, \emph{Phys. Rev. Lett.} \textbf{88}, 017901 (2001).

\bibitem{JPA_2001_Henderson} L. Henderson and V. Vedral, Classical, quantum and total correlations, \emph{J. Phys. A: Math. Gen.} \textbf{34}, 6899 (2001).

\bibitem{PRL_1998_Knill} E. Knill and R. Laflamme, Power of One Bit of Quantum Information, \emph{Phys. Rev. Lett.} \textbf{81}, 5672 (1998).

\bibitem{review_2012_Modi} K. Modi, A. Brodutch, H. Cable, T. Paterek, and V. Vedral, The classical-quantum boundary for correlations: Discord and related measures, \emph{Rev. Mod. Phys.} \textbf{84}, 1655 (2012).

\bibitem{review_2018_Bera} A. Bera, T. Das, D. Sadhukhan, S. S. Roy, A. Sen(De), and U. Sen, Quantum discord and its allies: a review of recent progress, \emph{Rep. Prog. Phys.} \textbf{81}, 024001 (2018).

\bibitem{PRA_2010_Soares} D. O. Soares-Pinto, L. C. C$\rm \acute{e}$leri, R. Auccaise, F. F. Fanchini, E. R. deAzevedo, J. Maziero, T. J. Bonagamba, and R. M. Serra, Nonclassical correlation in NMR quadrupolar systems, \emph{Phys. Rev. A} \textbf{81}, 062118 (2010).

\bibitem{PRL_2011_Auccaise} R. Auccaise, L. C. C$\rm \acute{e}$leri, D. O. Soares-Pinto, E. R. deAzevedo, J. Maziero, A. M. Souza, T. J. Bonagamba, R. S. Sarthour, I. S. Oliveira, and R. M. Serra, Environment-induced sudden transition in quantum discord dynamics, \emph{Phys. Rev. Lett} \textbf{107}, 140403 (2011).

\bibitem{PRA_2017_Singh} A. Singh, Arvind, and K. Dorai, Witnessing nonclassical correlations via a single-shot experiment on an ensemble of spins using nuclear magnetic resonance, \emph{Phys. Rev. A} \textbf{95}, 062318 (2017).

\bibitem{PRL_2008_Lanyon} B. P. Lanyon, M. Barbieri, M. P. Almeida, and A. G. White, Experimental quantum computing without entanglement, \emph{Phys. Rev. Lett.} \textbf{101}, 200501 (2008).

\bibitem{NC_2010_Xu} J. S. Xu, X. Y. Xu, C. F. Li, C. J. Zhang, X. B. Zou, and G. C. Guo, Experimental investigation of classical and quantum correlations under decoherence, \emph{Nat. Commun.} \textbf{1}, 7 (2010).

\bibitem{NC_2013_Xu}J.-S. Xu, K. Sun, C.-F. Li, X.-Y. Xu, G.-C. Guo, E. Andersson, R. Lo Franco, and G. Compagno, Experimental recovery of quantum correlations in absence of system-environment back-action, \emph{Nat. Commun.} \textbf{4}, 2851 (2013).

\bibitem{PRL_2015_Bromley} T. R. Bromley, M. Cianciaruso, and G. Adesso, Experimental entanglement activation from discord in a programmable quantum measurement, \emph{Phys. Rev. Lett.} \textbf{114}, 210401 (2015).

\bibitem{PRA_2016_Knoll} L. T. Knoll, C. T. Schmiegelow, O. J. Farias, S. P. Walborn, and M. A. Larotonda, Entanglement-breaking channels and entanglement sudden death, \emph{Phys. Rev. A} \textbf{94}, 012345 (2016).

\bibitem{PRB_2011_Yurishchev} M. A. Yurishchev, Quantum discord in spin-cluster materials, \emph{Phys. Rev. B} \textbf{84}, 024418 (2011).

\bibitem{PRB_2012_Rong} X. Rong, Z. Wang, F. Jin, J. Geng, P. Feng, N. Xu, Y. Wang, C. Ju, M. Shi, and J. Du, Quantum discord for investigating quantum correlations without entanglement in solids, \emph{Phys. Rev. B} \textbf{86}, 104425 (2012).

\bibitem{PRB_2013_Rong} X. Rong, F. Jin, Z. Wang, J. Geng, C. Ju, Y. Wang, R. Zhang, C. Duan, M. Shi, and J. Du, Experimental protection and revival of quantum correlation in open solid systems, \emph{Phys. Rev. B} \textbf{88}, 054419 (2013).

\bibitem{NP_2013_Gessner} M. Gessner, M. Ramm, T. Pruttivarasin, A. Buchleitner, H-P. Breuer, and H. H$\rm \ddot{a}$ffner,  Local detection of quantum correlations with a single trapped ion, \emph{Nat. Phys.} \textbf{10}, 105 (2014).

\bibitem{NC_2017_Abdelrahman} A. Abdelrahman, O. Khosravani, M. Gessner, A. Buchleitner, H. P. Breuer, D. Gorman, R. Masuda, T. Pruttivarasin, M. Ramm, P. Schindler and H. H$\rm \ddot{a}$ffner, Local probe of single phonon dynamics in warm ion crystals, \emph{Nat. Commun.} \textbf{8}, 15712 (2017).

\bibitem{PRA_2014_Wood} C. J. Wood, M. O. Abutaleb, M. G. Huber, M. Arif, D. G. Cory, and D. A. Pushin, Quantum correlations in a noisy neutron interferometer, \emph{Phys. Rev. A} \textbf{90}, 032315 (2014).

\bibitem{AQT_2019_Cozzolino} D. Cozzolino, B. Da Lio, D. Bacco, and L. K. Oxenlwe, High-dimensional quantum communication: benefits, progress, and future challenges, \emph{Adv. Quantum Technol.} \textbf{2}, 1900038 (2019).

\bibitem{FP_2020_Wang} Y. Wang, Z. Hu, B. C. Sanders and S. Kais, Qudits and high-dimensional quantum computing, \emph{Front. Phys.} \textbf{8}, 479 (2020).

\bibitem{review_2020_Erhard}M. Erhard, M. Krenn, and A. Zeilinger, Advances in high dimensional quantum entanglement, \emph{Nat. Rev. Phys.} \textbf{2}, 365 (2020).

\bibitem{PRL_2019_Wang} H. Wang, J. Qin, X. Ding, M.-C. Chen, S. Chen, X. You, Y.-M. He, X. Jiang, L. You, Z. Wang, C. Schneider, J.J. Renema, S. H${\rm \ddot{o}}$fling, C.-Y. Lu, and J.-W. Pan, Boson sampling with 20 input photons and a 60-Mode interferometer in a $10^{14}$-dimensional Hilbert space, \emph{Phys. Rev. Lett.} \textbf{123}, 250503 (2019).

\bibitem{NP_2019_Reimer} C. Reimer, S. Sciara, P. Roztocki, M. Islam, L. Romero Cort${\rm \acute{e}}$s, Y. Zhang, B. Fisher, S. Loranger, R. Kashyap, A. Cino, S. T. Chu, B.E. Little, D.J. Moss, L. Caspani, W.J. Munro, J. Aza${\rm \tilde{n}}$a, M. Kues, and R. Morandotti, High-dimensional one-way quantum processing implemented on d-level cluster states, \emph{Nat. Phys.} \textbf{15}, 148-153 (2019).

\bibitem{PRL_2019_Luo} Y.-H. Luo, H.-S. Zhong, M. Erhard, X.-L. Wang, L.-C. Peng, M. Krenn, X. Jiang, L. Li, N.-L. Liu, C.-Y. Lu, A. Zeilinger, J.-W. Pan, Quantum Teleportation in High Dimensions, \emph{Phys. Rev. Lett.} \textbf{123}, 070505 (2019).

\bibitem{PRL_2020_Hu} X.-M. Hu, C. Zhang, B.-H. Liu, Y. Cai, X.-J. Ye, Y. Guo, W.-B. Xing, C.-X. Huang, Y.-F. Huang, C.-F. Li, and G.-C. Guo, Experimental high-dimensional quantum teleportation, \emph{Phys. Rev. Lett.} \textbf{125}, 230501 (2020).

\bibitem{PRL_2010_Vertesi} T. V$\acute{\rm e}$rtesi, S. Pironio, and N. Brunner, Closing the detection loophole in bell experiments using qudits, \emph{Phys. Rev. Lett.} \textbf{104}, 060401 (2010).

\bibitem{nature_2017_Kues} M. Kues, C. Reimer, P. Roztocki, L.R. Cort$\rm \acute{e}$s, S. Sciara, B. Wetzel, Y. Zhang, A. Cino, S.T. Chu, B.E. Little, D. J. Moss, L. Caspani, J. Aza$\rm \tilde{n}$a, and R. Morandotti, On-chip generation of high-dimensional entangled quantum states and their coherent control, \emph{Nature} \textbf{546},  622 (2017).

\bibitem{Science_2018_Wang} J. Wang, S. Paesani, Y. Ding, R. Santagati, P. Skrzypczyk, A. Salavrakos, J. Tura, R. Augusiak, L. Man$\rm \check{c}$inska, D. Bacco, D. Bonneau, J.W. Silverstone, Q. Gong, A. Ac$\rm \acute{i}$n, K. Rottwitt, L.K. Oxenl\o{}we, J.L. O'Brien, A. Laing, and M.G. Thompson, Multidimensional quantum entanglement with large-scale integrated optics, \emph{Science} \textbf{360}, 285 (2018).

\bibitem{NPJ_2020_Lu}L. Lu, L. Xia, Z. Chen, L. Chen, T. Yu, T. Tao, W. Ma, Y. Pan, X. Cai, Y. Lu, S. Zhu, and X.-S. Ma, Three-dimensional entanglement on a silicon chip, \emph{Npj Quantum Inf.} \textbf{6}, 30 (2020).

\bibitem{arxiv_2021_Ding}A. Cervera-Lierta, M. Krenn, A. Aspuru-Guzik, and A. Galda, Experimental high-dimensional Greenberger-Horne-Zeilinger entanglement with superconducting transmon qutrits, \emph{Phys. Rev. Applied} \textbf{17}, 024062 (2022).

\bibitem{LSA_2016_Ding}D.-S. Ding, W. Zhang, S. Shi, Z.-Y. Zhou, Y. Li, B.-S. Shi, and G.-C. Guo, High-dimensional entanglement between distant atomic-ensemble memories, \emph{Light: Sci. Appl.} \textbf{5}, e16157 (2016).

\bibitem{PRL_2018_Zeng} Q. Zeng, B. Wang, P.-Y. Li, and X.-D. Zhang, Experimental high-dimensional Einstein-Podolsky-Rosen steering, \emph{Phys. Rev. Lett.} \textbf{120}, 030401 (2018).

\bibitem{PRL_2019_Guo} Y. Guo, S. Cheng, X. Hu, B.-H. Liu, E.-M. Huang, Y.-F.Huang, C.-F. Li, G.-C. Guo, and E. G. Cavalcanti, Experimental measurement-device-independent quantum steering and randomness generation beyond qubits, \emph{Phys. Rev. Lett.} \textbf{123}, 170402 (2019).

\bibitem{PRL_2021_Designolle} S. Designolle, V. Srivastav, R. Uola, N. H. Valencia, W. McCutcheon, M. Malik, and N. Brunner, Genuine high-dimensional quantum steering, \emph{Phys. Rev. Lett.} \textbf{126}, 200404 (2021).

\bibitem{PhysRep_2013_Doherty} M. W. Doherty, N. B. Manson, P. Delaney, F. Jelezko, J. Wrachtrup, and L. C. L. Hollenberg, The nitrogen-vacancy colour centre in diamond, \emph{Phys. Rep.} \textbf{528}, 1 (2013).

\bibitem{QIP_2013_Ye} B. Ye, Y. Liu, J. Chen, X. Liu, and Z. Zhang,  Analytic expressions of quantum correlations in qutrit Werner states, \emph{Quantum Inf. Proc.} \textbf{12}, 2355 (2013).

\bibitem{PRA_2006_Baumgartner} B. Baumgartner, B. C. Hiesmayr, and H. Narnhofer, State space for two qutrits has a phase space structure in its core, \emph{Phys. Rev. A} \textbf{74},  032327 (2006).

\bibitem{PRA_2012_Chitambar} E. Chitambar, Quantum correlations in high-dimensional states of high symmetry, \emph{Phys. Rev. A} \textbf{86}, 032110 (2012).

\bibitem{PRA_2021_Poxleitner} M. Poxleitner, H. Hinrichsen, Gaussian continuous-variable isotropic state, \emph{Phys. Rev. A} \textbf{104}, 032423 (2021).

\bibitem{PRL_2009_Jacques} V. Jacques, P. Neumann, J. Beck, M. Markham, D. Twitchen, J. Meijer, F. Kaiser, G. Balasubramanian, F. Jelezko, and J. Wrachtrup, Dynamic polarization of single nuclear spins by optical pumping of nitrogen-vacancy color centers in diamond at room temperature, \emph{Phys. Rev. Lett.} \textbf{102}, 057403 (2009).

\bibitem{PRB_2013_Fischer} R. Fischer, A. Jarmola, P. Kehayias, and D. Budker, Optical polarization of nuclear ensembles in diamond, \emph{Phys. Rev. B} \textbf{87}, 125207 (2013).

\bibitem{PRA_2001_Daniel} D. F. James, P. G. Kwiat, W. J. Munro, and A. G. White, Measurement of qubits, \emph{Phys. Rev. A} \textbf{64}, 052312 (2001).

\bibitem{SM} See Supplemental Material for details of the state preparation and reconstruction, experimental procedures, and the calculation of the quantum discord and entanglement, which includes Ref. \cite{nature_2012_Van}.

\bibitem{nature_2012_Van}T. Van der Sar, Z. Wang, M. Blok, H. Bernien, T. Taminiau, D. Toyli, D. Lidar, D. Awschalom, R. Hanson, and V. Dobrovitski, Decoherence-protected quantum gates for a hybrid solid-state spin register, \emph{Nature} \textbf{484}, 82-86 (2012).

\bibitem{PRA_2012_Rossignoli} R. Rossignoli, J. M. Matera, and N. Canosa, Measurements, quantum discord, and parity in spin-1 systems, \emph{Phys. Rev. A 86} \textbf{473}, 022104 (2012).

\bibitem{PRA_2007_Derkacz}L. Derkacz and L. Jakobczyk, Entanglement versus entropy for a class of mixed two-qutrit states, \emph{Phys. Rev. A} \textbf{76}, 042304 (2007).

\bibitem{TCS_2004_Biham} E. Biham, G. Brassard, D. Kenigsberg, and T. Mora, Quantum computing without entanglement, \emph{Theor. Comput. Sci.} \textbf{320}, 15 (2004).

\bibitem{science_2007_Almeida} M.P. Almeida, F. de Melo, M. Hor-Meyll, A. Salles, S.P. Walborn, P.H.S. Ribeiro, and L. Davidovich, Environment-induced sudden death of entanglement, \emph{Science} \textbf{316}, 579 (2007).

\bibitem{PRL_2021_Yurtalan}M. A. Yurtalan, J. Shi, M. Kononenko, A. Lupascu, and S. Ashhab, Implementation of a Walsh-Hadamard gate in a superconducting qutrit, \emph{Phys. Rev. Lett.} \textbf{125}, 180504 (2020).

\bibitem{PRX_2021}M. Blok, V. Ramasesh, T. Schuster, K. OBrien, J. Kreikebaum, D. Dahlen, A. Morvan, B. Yoshida, N. Yao, and I. Siddiqi, Quantum information scrambling on a superconducting qutrit processor, \emph{Phys. Rev. X} \textbf{11}, 021010 (2021).

\bibitem{PRL_2021_Morvan}A. Morvan, V. V. Ramasesh, M. S. Blok, J. M. Kreikebaum, K. O'Brien, L. Chen, B. K. Mitchell, R. K. Naik, D. I. Santiago, and I. Siddiqi, Qutrit randomized Benchmarking, \emph{Phys. Rev. Lett.} \textbf{126}, 210504 (2021).

\bibitem{science_2021} M. Pompili, S. L. N. Hermans, S. Baier, H. K. C. Beukers, P. C. Humphreys, R. N. Schouten, R. F. L. Vermeulen, M. J. Tiggelman, L. dos Santos Martins, and B. Dirkse, Realization of a multinode quantum network of remote solid-state qubits, \emph{Science} \textbf{372}, 259 (2021).

\bibitem{Nphys_2013}F. Dolde, I. Jakobi, B. Naydenov, N. Zhao, S. Pezzagna, C. Trautmann, J. Meijer, P. Neumann, F. Jelezko, and J. Wrachtrup, Room-temperature entanglement between single defect spins in diamond, \emph{Nat. Phys.} \textbf{9}, 139 (2013).

\bibitem{NC_2014}F. Dolde, V. Bergholm, Y. Wang, I. Jakobi, B. Naydenov, S. Pezzagna, J. Meijer, F. Jelezko, P. Neumann, T. Schulte-Herbr$\ddot{u}$ggen J. Biamonte, and J. Wrachtrup, High-fidelity spin entanglement using optimal control, \emph{Nat. Commun.} \textbf{5}, 3371 (2014).

\bibitem{nature_2022} S. L. N. Hermans, M. Pompili, H. K. C. Beukers, S. Baier, J. Borregaard, and R. Hanson, Qubit teleportation between non-neighbouring nodes in a quantum network, \emph{Nature} \textbf{605}, 663-668 (2022).



\end{thebibliography}

\begin{thebibliography}{99}
\setcounter{enumiv}{0}

\bibitem{PRL_V.Jacques} V. Jacques et al, Dynamic Polarization of Single Nuclear Spins by Optical Pumping of Nitrogen-Vacancy Color Centers in Diamond at Room Temperature, \emph{Phys. Rev. Lett.} \textbf{102}, 057403 (2009).

\bibitem{PRA_2001_Daniel} D. F. James, and  P. G. Kwiat, W. J. Munro, and A. G. White, Measurement of qubits, \emph{Phys. Rev. A} \textbf{64}, 052312 (2001).

\bibitem{review_2018_Bera} A. Bera, T. Das, D. Sadhukhan, S. S. Roy, A. Sen(De), and U. Sen, Quantum discord and its allies: a review of recent progress, \emph{Rep. Prog. Phys.} \textbf{81}, 024001 (2018).

\bibitem{PRA_2012_Rossignoli} R. Rossignoli, J. M. Matera, and N. Canosa, Measurements, quantum discord, and parity in spin-1 systems,  \emph{Phys. Rev. A } \textbf{86}, 022104 (2012).

\bibitem{PRA_2007_Derkacz}  {\L}. Derkacz and L. Jak\'{o}bczyk, Entanglement versus entropy for a class of mixed two-qutrit states, \emph{ Phys. Rev. A} \textbf{76}, 042304 (2007).

\bibitem{Nature_2012_Sar}  T.V. Sar et al, Decoherence-protected quantum gates for a hybrid solid-state spin register, \emph{Nature} \textbf{484}, 82-86 (2012).

\end{thebibliography}
\end{document}